\documentclass{jfm}
\usepackage[usenames,dvipsnames]{xcolor}

\usepackage{charter}
\usepackage{amssymb}
\usepackage{amsmath}
\usepackage{natbib}
\usepackage{epsfig}
\usepackage{psfrag}
\usepackage{marginnote}
\usepackage[compact]{titlesec}
\usepackage[a4paper,ps2pdf,colorlinks=true,frenchlinks=true,breaklinks=true]{hyperref}
\usepackage{soul}

\allowdisplaybreaks
\numberwithin{equation}{section}

\ifCUPmtlplainloaded \else
  \checkfont{eurm10}
  \iffontfound
    \IfFileExists{upmath.sty}
      {\typeout{^^JFound AMS Euler Roman fonts on the system,
                   using the 'upmath' package.^^J}%
       \usepackage{upmath}}
      {\typeout{^^JFound AMS Euler Roman fonts on the system, but you
                   dont seem to have the}%
       \typeout{'upmath' package installed. JFM.cls can take advantage
                 of these fonts,^^Jif you use 'upmath' package.^^J}%
      }
  \else
  \fi
\fi

\ifCUPmtlplainloaded \else
  \checkfont{msam10}
  \iffontfound
    \IfFileExists{amssymb.sty}
      {\typeout{^^JFound AMS Symbol fonts on the system, using the
                'amssymb' package.^^J}%
       \usepackage{amssymb}%

      }{}
  \fi
\fi

\ifCUPmtlplainloaded \else
  \IfFileExists{amsbsy.sty}
    {\typeout{^^JFound the 'amsbsy' package on the system, using it.^^J}%
     \usepackage{amsbsy}}
    {\providecommand\boldsymbol[1]{\mbox{\boldmath $##1$}}}
\fi

\providecommand\bnabla{\boldsymbol{\nabla}}

\newsavebox{\astrutbox}
\sbox{\astrutbox}{\rule[-5pt]{0pt}{20pt}}

\renewcommand{\vec}[1]{\mbox{\boldmath $ #1$}}

\newcommand{\R}{\ensuremath{{\mathit{R}}}}
\newcommand{\Ro}{\ensuremath{{\mathit{R}}}}
\newcommand{\omo}{\ensuremath{{\mathit{\omega}}}}
\newcommand{\Rc}{\ensuremath{\mathit{R_c}}}
\newcommand{\alpc}{\ensuremath{\mathit{\alpha_c}}}
\newcommand{\Ra}{\ensuremath{{\mathit{Ra}}}}
\newcommand{\Rac}{\ensuremath{{\mathit{Ra_c}}}}
\renewcommand{\Pr}{\ensuremath{{\mathit{Pr}}}}
\newcommand{\etaP}{\ensuremath{{\mathit{\eta_P}}}}

\newcommand{\red}[1]{#1}
\newcommand{\blue}[1]{#1}
\newcommand{\brown}[1]{#1}
\newcommand{\reviii}[1]{#1}
\newcommand{\reviv}[1]{#1}

\title[Quasi-geostrophic approximation of anelastic convection]{Quasi-geostrophic approximation of anelastic convection}
\author[F.~H.~Busse and R.~D.~Simitev]%
{Friedrich H.~Busse$^{1,3,}$\thanks{Email address for correspondence: busse@uni-bayreuth.de}
and Radostin D.~Simitev$^{2,3,4}$}

\affiliation{%
$^1$Institute of Physics, University of Bayreuth, Bayreuth D-95440, Germany\\
$^2$School of Mathematics and Statistics, University of Glasgow, Glasgow G12 8QW, UK \\
$^3$ Earth and Space Sciences, University of California Los Angeles, Los Angeles CA 90095, USA\\
$^4$ Hansen Experimental Physics Laboratory, Stanford University, Stanford CA 94305, USA}

\pubyear{2014}
\volume{751}
\pagerange{216--217}
\date{28 July 2013; revised 19 May 2014; accepted 22 May 2014}
\doi{jfm.2014.293}
\begin{document}

\maketitle

\begin{abstract}
The onset of convection in a rotating cylindrical annulus with
parallel ends filled with a compressible fluid  is studied in the
anelastic approximation. Thermal Rossby waves propagating in the
azimuthal direction are found as solutions. The analogy to the case of
Boussinesq convection in the presence of conical end surfaces of the
annular region is emphasised. As in the latter case the results can be
applied as an approximation for the description of the onset of
anelastic convection in rotating spherical fluid shells. Reasonable
agreement with three-dimensional numerical results published \blue{by
Jones {\em et~al.} ({\em J.~Fluid Mech.}, vol.~634, 2009, pp.~291-319)}
for the latter problem is found.
\red{As in those results the location of the onset of convection shifts
outwards from the tangent cylinder with increasing number $N_\rho$ of
density scale heights until it reaches the equatorial boundary. A new
result is that at a much higher number $N_\rho$ the onset location returns to
the interior of the fluid shell.}

\end{abstract}


\begin{figure}
\begin{center}
\vspace{0mm}
\begin{tabular}{ll}
(a) & (b)\\
\raisebox{0mm}{\epsfig{file=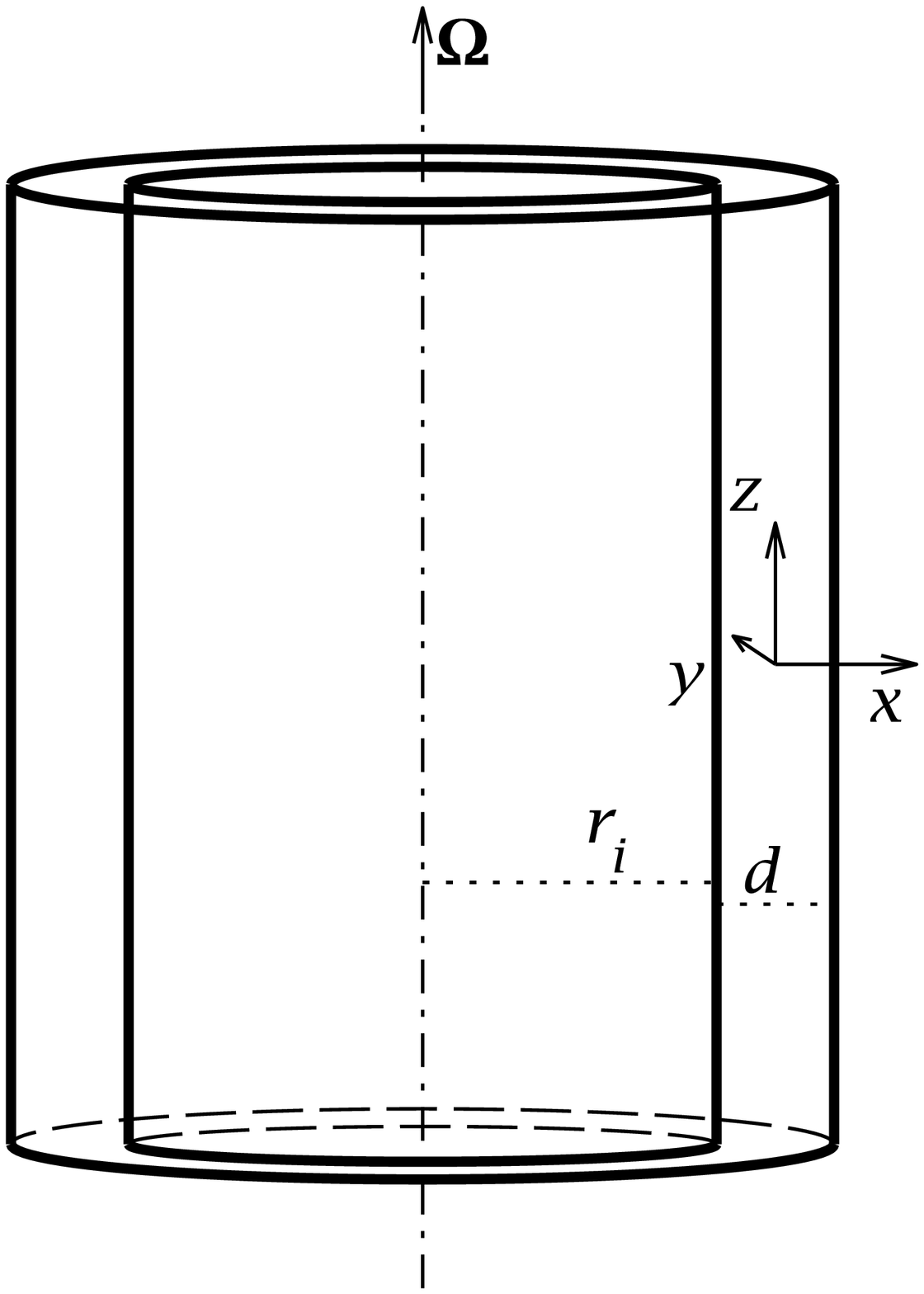,width=0.46\textwidth,clip=}
} &
\raisebox{0mm}{\epsfig{file=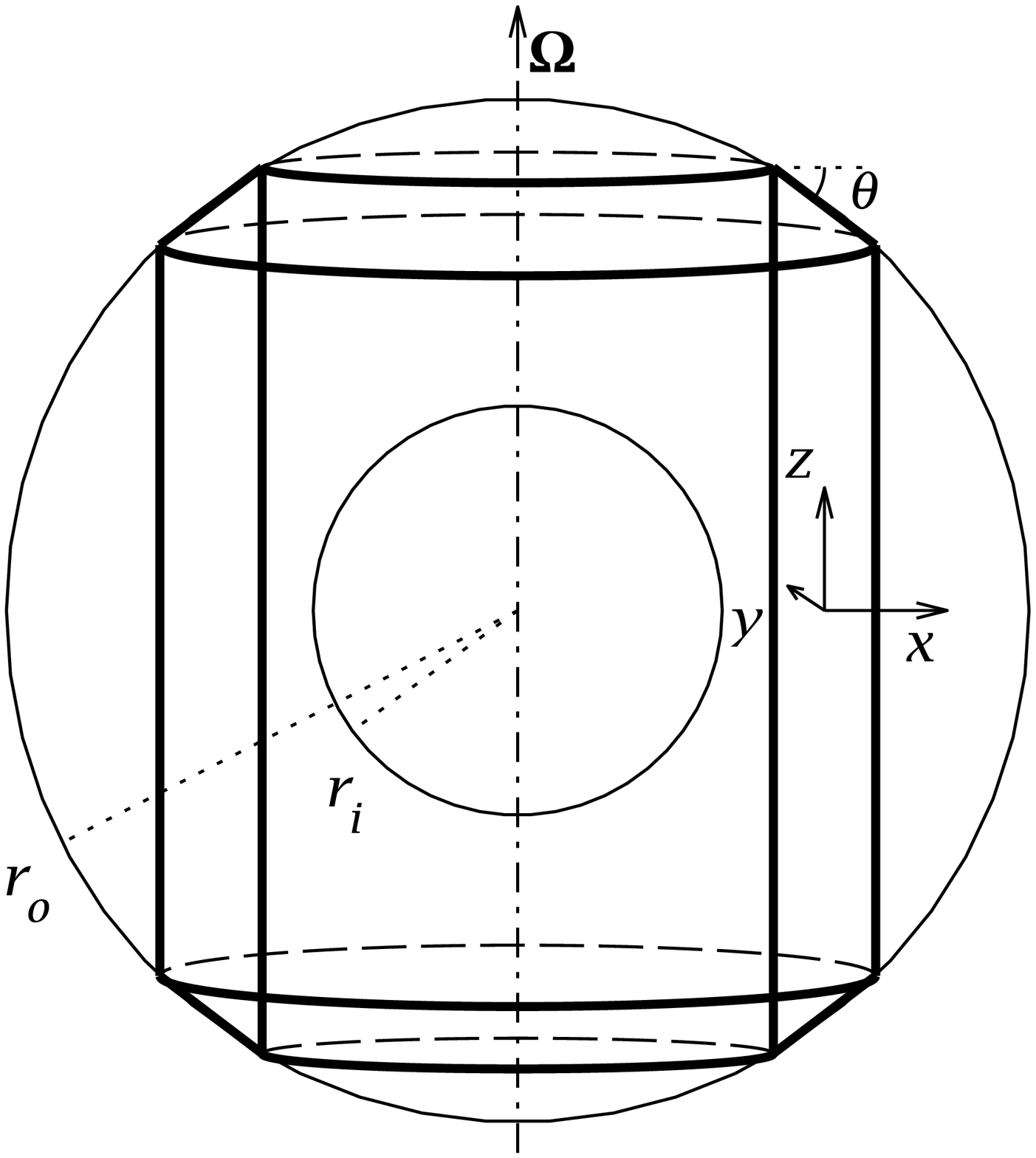,width=0.46\textwidth,clip=}
} \\
\end{tabular}
\end{center}
\caption[]{Sketch of geometric configurations. (a) The rotating
  cylindrical annulus with parallel ends considered in \textsection
  \ref{sec2} as a model of the effect of background density
  variation. (b) The rotating cylindrical annulus with conical
  ends inscribed in a spherical shell. 
  The sketches are not to scale with the asymptotic limit assumptions.  
}
\label{f.010}
\end{figure}

\section{Introduction}
\label{sec1}

The tendency of fluid motions in rapidly rotating systems to develop
nearly two-dimensional structures has often been exploited to simplify
the theoretical analysis. The description of convection flows in
systems where the gravity vector and the rotation axis are not parallel
provides a typical example \citep{Bu70,Bu02}. In applications of
convection problems to rotating planets and stars the tendency towards
two-dimensionality is partly obscured by the strong variation of fluid
density as a function of radius in the nearly spherical systems. It is
thus of interest to investigate the extent to which the
quasi-geostrophic
two-dimensional description can still provide an approximation for
three-dimensional convection in rapidly rotating systems with strong
variations of density. 

In the case of the Boussinesq approximation, in which the density is
regarded as constant except in connection with the gravity term, the
results derived from the two-dimensional \red{quasi-geostrophic} analysis of the
onset of convection in rotating spherical fluid shells \red{compare} well
with the results of the three-dimensional numerical analysis
\citep{Simitev03}. In this \red{paper} the two-dimensional model was
based on the problem of convection in a rotating cylindrical 
annulus with conical end boundaries \citep{Bu70, Bu86}. 
\reviv{For a more detailed} {discussion of the role of the quasi-geostrophic model in
  relationship to more accurate three-dimensional solutions for
  convection in rotating spherical fluid shells we refer} \red{to the paper
  of \cite{Gi06}.} 

In recent years the anelastic approximation \citep{Go69} has been
widely used to obtain more realistic descriptions of convection in the
atmospheres of planets and stars with strong variations of density. In
the paper by \citet[hereafter B86]{Bu86}, the analogy between the effect of changing height induced
by the conical boundaries of the cylindrical annulus and the effect of a
radial variation of density \reviv{has} already been pointed out. In the
present paper we intend to demonstrate quantitatively that the
two-dimensional analysis of the annulus model provides a
reasonable approximation for the onset of convection in the presence
of strong anelastic density variations in rotating spherical fluid
shells.

The analysis of \blue{the present} paper resembles to some extent the
two-dimensional analysis of anelastic convection pursued by  
\citet{Evo04,Evo06}; see also \citet{Gla09}. Because of the
high computational cost of three-dimensional simulations of convection
in the presence of density variation over many scale heights these
authors restricted their attention to two-dimensional numerical simulation of
convection close to the equatorial plane. Evonuk \& Glatzmaier were
interested in the nonlinear properties of two-dimensional convection
including zonal flows in the presence of strong density variations. In
contrast, our analytical model focuses on the linear problem of the
onset of convection at high values of the rotation parameter. 

The main purpose of this paper is not the demonstration of a high
accuracy of the two-dimensional  approximation. Instead we wish to
emphasise the insights into anelastic convection in rotating spheres
gained from the analytical quasi-geostrophic model.
In the \red{next section} we first introduce the narrow-gap
cylindrical annulus with parallel ends as shown in figure
\ref{f.010}(a) and derive the two-dimensional solution describing
anelastic convection. In \textsection \ref{sec3} the model is modified for
applications to the onset of anelastic convection in rotating
spherical shells as indicated in figure \ref{f.010}(b). \blue{Detailed}
comparisons with numerical solutions are evaluated in \textsection \ref{sec4}. Some
\reviv{nonlinear aspects} are discussed in the final \textsection of the paper.      

\section{Mathematical description of two-dimensional anelastic convection}
\label{sec2}

We consider a cylindrical annulus with parallel ends rotating about its axis with the
angular velocity $\vec \Omega$ as shown in figure \ref{f.010}(a). The gap
width $d$ in the radial direction of the annular region is small in
comparison with its inner radius $r_i$ such that a cartesian  system
of dimensionless coordinates $x, y, z$ in the radial, azimuthal and
axial directions, respectively, can be used for a local description of
convection. The corresponding unit vectors are $\vec i, \vec j$ and
$\vec k$. The annular gap is filled with an ideal gas the state of
which differs little from an isentropic reference state in the
presence of gravity pointing in the negative $x$-direction. The small
deviation from the isentropic state is described by the small positive
excess entropy $\Delta s$ by which the entropy at the inner cylinder
exceeds the entropy at the outer cylindrical boundary. In experimental
realisations gravity could be replaced by the centrifugal force. The
dynamical problem would then be identical \reviv{if} a negative value of
$\Delta s$ is assumed. 
 
Using $d$ as length scale, $d^2/\kappa$ as time scale and $\Delta s$
as the scale of the entropy we obtain the dimensionless form of the
\red{anelastic equations as used in \citep{Jones09} and in the
benchmark paper \citep{Jones11} based on the formulation introduced independently by \citet{La99} and \citet{Br95},} 
\begin{subequations}
\label{eq:1}
\begin{gather}
\label{eq:1a}
\frac{\partial\vec u}{\partial t}+\vec u\cdot \bnabla\vec u+ \tau\vec
k\times \vec u  =  -\bnabla \pi \red{+} \vec i \frac{\R}{\Pr} s + \vec F,\\ 
\label{eq:1b}
\bnabla\cdot \vec u = \bar\rho\vec u\cdot \bnabla \frac{1}{\bar\rho},\\
\label{eq:1c}
\Pr\left(\frac{\partial s}{\partial t} +\vec u\cdot\bnabla s\right) = \bnabla^2 s+\bnabla s\cdot\frac{1}{\kappa\bar\rho\overline T} \bnabla
\kappa\bar\rho\overline T +\hat Q,
\end{gather}
\end{subequations}
where $\vec F$ denotes the force of viscous friction divided by the
density, and all terms in the equation of motion that can be written
as gradients have been combined into $\bnabla \pi$.
The Rayleigh number $\R$, the Prandtl number $\Pr$ and the Coriolis number
$\tau$ are defined by  
\begin{gather}
\R=\frac{gd^3\Delta s}{\kappa\nu c_p}, \hspace{1.0cm}
 \Pr=\frac{\nu}{\kappa},\hspace{1.0cm} \tau = 2\Omega \frac{d^2}{\nu}.
\label{eq:2}
\end{gather}
Here $\kappa$ is the entropy diffusivity, $\nu$ is the kinematic
viscosity and $c_p$ is the specific heat at constant pressure. For
simplicity we have assumed that material properties 
are constant except for $\bar\rho(x)$, which represents the $x$-dependent
density of the isentropic reference state made dimensionless through
division by its average value, and $\overline T(x)$ is the temperature
profile of that state. The constant gravity vector is given by $\vec
g= - g\vec i$, and the entropy $s$ can be separated into two parts,  
\begin{gather}
 s = -x + \tilde s,
\label{eq:3}
\end{gather}
such that the boundary condition $\tilde s =0$ holds at $x=\pm 1/2$.

We now consider two-dimensional solutions of equations \eqref{eq:1}
that are 
independent of $z$ and thus satisfy the Proudman-Taylor condition.
Assuming $\vec u= (1/\bar\rho)\bnabla\psi (x,y,t)\times \vec k$ we
obtain for the $z$ component of the vorticity of equation \eqref{eq:1a}
\begin{gather}
\frac{\partial \zeta}{\partial
  t}+\frac{1}{\bar\rho}\left(\frac{\partial \zeta}{\partial x}
\frac{\partial\psi}{\partial y} - \frac{\partial\zeta}{\partial
  y}\frac{\partial\psi}{\partial x} - 
(\tau + \zeta)\frac{1}{\bar\rho}\frac{d
  \bar\rho}{dx}\frac{\partial\psi}{\partial y}\right) =   -
\frac{\R}{\Pr}\frac{\partial \tilde s}{\partial y} + \Delta_2\zeta,
\label{eq:4}
\end{gather}
where $\zeta=\vec k\cdot \bnabla\times\big((\bnabla\psi\times\vec
k)/\bar\rho\big)$ is the $z$ component of the vorticity and
$\partial^2/\partial x^2 + \partial^2/\partial y^2$ has been denoted
by $\Delta_2$. Following \citet{Evo04} the friction
term has been reduced to its main contributor. 
Assuming that $\bar\rho$ varies slowly such that the absolute value of
\begin{gather}
\eta_\rho\equiv-\frac{1}{\bar\rho}\frac{d\bar\rho}{dx}
\label{eq:5}
\end{gather}
is a small constant we find that the absolute value of 
\begin{gather}
\eta_\rho^*\equiv\eta_\rho\tau
\label{eq:6}
\end{gather}
is a parameter of the order unity or larger for $\tau \gg 1$.
\reviv{In fact, later we shall consider the limit of $\eta_\rho^*$
tending to infinity.} 

We thus have arrived at the same equations as in the case of Boussinesq
convection in the annulus with conical end boundaries (see figure
\ref{f.010}(b)) given by (4.1) of B86 with the only  difference that the
second and third terms on the right-hand side of equation \eqref{eq:1c} are
missing in the latter case. In 
the following we shall neglect these two terms anticipating that they
become negligible in the asymptotic solution of the problem \red{for $\tau$ tending to infinity.}

\begin{figure}
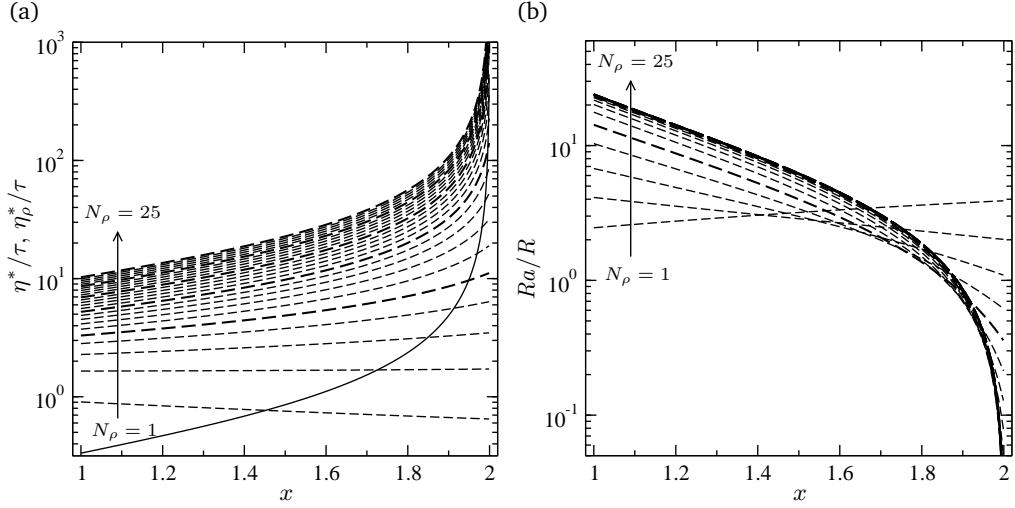

\begin{center}
\psfrag{eta}{\hspace{-4mm}$\eta^*/\tau, \, \eta_\rho^*/\tau$}
\psfrag{RaR}{\hspace{-4mm}$\Ra/\R$}
\psfrag{Ra}{$\Rac$}
\psfrag{Nr=1}{{\scriptsize \hspace*{-1.5mm} $N_\rho=1$}}
\psfrag{Nr=25}{{\scriptsize \hspace*{-2.5mm} $N_\rho=25$}}
\psfrag{x}{$x$}
\psfrag{(a)}{}
\psfrag{(b)}{}
\begin{tabular}{ll}
(a) & (b) \\
\epsfig{file=fig02a.eps,width=0.485\textwidth,clip=} &
\epsfig{file=fig02b.eps,width=0.491\textwidth,clip=} \\
\end{tabular}
\end{center}
\blue{\caption{{(a) The functions $\eta^*/\tau$ given by \eqref{eq:14}
    (solid line) and $\eta_\rho^*/\tau$ given by \eqref{eq:16} (dashed
    lines), and (b) the factor $\Ra/\R$ given by \eqref{eq:18} in
    dependence of $x$ in the case $n=2$,
    $\beta=0.5$ and $N_\rho = 1,2,\dotsc,25$ increasing in the
    direction of the arrows with the thicker dashed  lines
    corresponding to $N_\rho = 5,10, 15, 20, 25$. 
\label{fig02a}
}}}
\end{figure}

The linearised versions of equations \eqref{eq:1c} and \eqref{eq:4} \red{thus} assume the forms
\begin{subequations}
\label{eq:7}
\begin{align}
\frac{\partial \zeta}{\partial t}+
\frac{\eta_\rho^*}{\bar\rho}\frac{\partial \psi}{\partial y}&=   -
\frac{\R}{\Pr}\frac{\partial \tilde s}{\partial y}+ \Delta_2\zeta,\\
\Pr\frac{\partial \tilde s}{\partial t} - \frac{\Pr}{\bar\rho}\frac{\partial\psi}{\partial y} &= \Delta_2\tilde s.
\end{align}
\end{subequations}
After elimination of $\tilde s$, neglecting terms of the order of $\eta_\rho$,
and multiplication of the equation of motion by $\bar\rho$ we obtain 
\begin{gather}
\left(\Pr\frac{\partial}{\partial t}-\Delta_2\right)\left[\Big(\frac{\partial}{\partial
      t}-\Delta_2\Big)\Delta_2 - \eta_\rho^*\frac{\partial}{\partial
      y}\right]\psi =  \R \frac{\partial^2}{\partial y^2}\psi.
\label{eq:8}
\end{gather}
This equation is easily solved when stress-free conditions at the
boundaries $x=\pm 1/2$ are assumed, 
\begin{gather}
\psi = \sin\big(\pi(x+1/2)\big)\exp(i\alpha y-i\omega t),  
\hspace{1.0cm} 
\tilde s=\frac{-i\alpha\psi}{-i\omega \Pr +\alpha^2 +\pi^2}.
\label{eq:9}
\end{gather}
This solution yields the dispersion relation
\begin{gather}
\Ro\alpha^2=(-i\omega \Pr +\alpha^2 +\pi^2)\big[(-i\omega  +\alpha^2 +\pi^2)(\alpha^2 +\pi^2)+i\alpha\eta_\rho^*\big].
\label{eq:10}
\end{gather}
The real and imaginary parts of this equation determine the neutral curve
$\Ro(\alpha)$ \reviv{and the frequency $\omo(\alpha)$} of the thermal Rossby wave,
\begin{gather}
\omo=\frac{\alpha\eta_\rho^*}{(1+\Pr)(\alpha^2 + \pi^2)},  
\hspace{0.5cm} 
\Ro= (\alpha^2 + \pi^2)^3\alpha^{-2}+\left(\frac{\eta_\rho^*\Pr}{1+\Pr}\right)^2/(\alpha^2 + \pi^2).
\label{eq:11}
\end{gather}
The angular frequency $\omo$ resembles that of ordinary Rossby
waves from which it differs only through the appearance of $\Pr$ in the
denominator. The Rayleigh number is determined by two terms. The first
is the familiar expression from Rayleigh-B\'enard convection which is
independent of the Coriolis number. The second term is introduced by
the density variation caused by the compressibility. 

The critical value $\Rc$ and the corresponding wavenumber $\alpc$
are obtained through minimising $\Ro(\alpha)$ with respect to
$\alpha$ which yields in the
limit of high values of $|\eta_\rho^*|$   
\begin{gather}
\alpc=\etaP^{1/3}(1-\frac{7}{12}\pi^2\etaP^{-2/3} + \dots),  
\hspace{1cm} 
\Rc= \etaP^{4/3}(3+\pi^2\etaP^{-2/3} + \dots),
\label{eq:12}
\end{gather}
where $\etaP$ is defined by
\begin{gather}
\etaP\equiv\frac{|\eta_\rho^*|\Pr}{\sqrt{2}(1+\Pr)}.
\label{eq:13}
\end{gather}
As in the Boussinesq case of the cylindrical annulus with conical
axial boundaries, the onset of convection becomes independent of the
gap width $d$ in the limit of high $|\eta_\rho^*|$ and the Rayleigh numbers
for modes with $\sin\big(l\pi(x+1/2)\big)$ with $l=2,3,4,\dotsc$ hardly differ
from that for $l=1$. The neglected second term on the right-hand side
of equation \eqref{eq:1c} would contribute only a negligible amount in the
limit of high $\alpc$. The other neglected term $\hat Q$ does not
enter the linear problem, of course. 

The fact that the asymptotic relationship \red{$\Rc=3\alpc^4$} holds
independently of the choice of length scale for a given entropy
gradient is essential for the following analysis.
\reviii{
In other words, the only remaining physical length scale in the limit
\begin{equation}
\eta_P^{1/3} \gg 1
\label{lenghtscale}
\end{equation}
is the dimensional azimuthal wave length $\lambda_D$ of convection
which must be small compared to any radial length scale of the
problem. In terms of dimensional quantities the condition
\eqref{lenghtscale} translates into the condition that the asymptotic
expression for $\lambda_D$,
\begin{equation}
\lambda_D \equiv 2\pi
\left(\frac{\kappa\sqrt{2}(1+\Pr)\overline{\rho}}{2\Omega \left(\displaystyle\frac{d}{d x_D}\overline{\rho}\right)}\right)^{1/3}
\label{lenghtscale1}
\end{equation}
is small compared to any radial length scale of the problem, where
$x_D$ denotes the dimensional length on
which $\overline{\rho}$ varies.
}
\begin{figure}
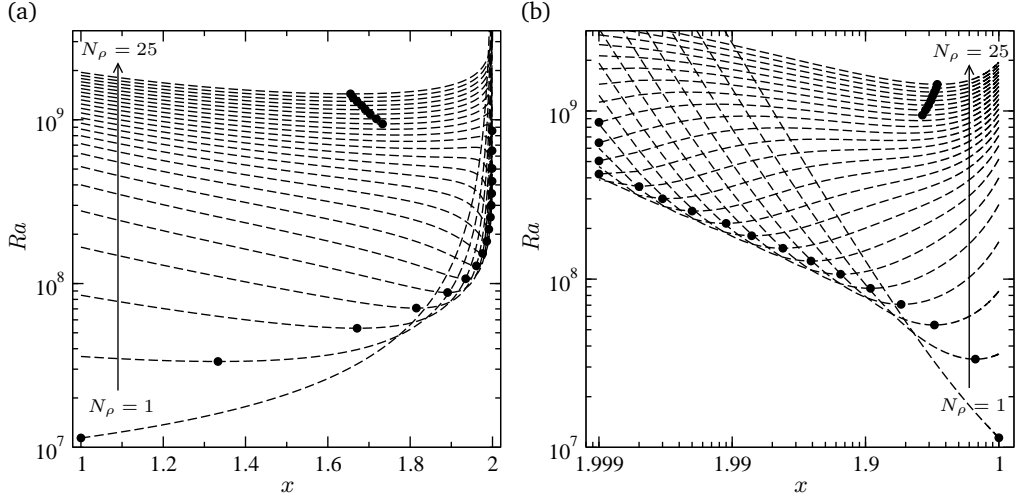

\begin{center}
\psfrag{Ra}{$\Ra$}
\psfrag{Nr=1}{{\scriptsize \hspace*{-1.5mm} $N_\rho=1$}}
\psfrag{Nr=25}{{\scriptsize \hspace*{-2.5mm} $N_\rho=25$}}
\psfrag{x}{$x$}
\psfrag{(a)}{}
\psfrag{(b)}{}
\begin{tabular}{ll}
(a) & (b)\\
\epsfig{file=fig03a.eps,width=0.49\textwidth,clip=} &
\epsfig{file=fig03b.eps,width=0.49\textwidth,clip=} \\
\end{tabular}
\end{center}
\caption{Critical Rayleigh number as a function of position $x$ in the cases
  $\beta=0.5$, $\Pr=1$, $\tau=10^5$, $n=2$ and $N_\rho = 1,2,\dotsc,25$
  increasing in the direction of the arrows.  The black dots
  indicate the position of the minimum on each curve. Panels (a) and
  (b) are identical only the  $x$-axis is scaled differently to reveal
  the structure near the outer boundary. }  
\label{fig02}
\end{figure}

\section{Application to three-dimensional geometries} 
\label{sec3}

In applying the two-dimensional solution to a three-dimensional
configuration we follow the corresponding analysis \reviv{\citep{Bu70}} in the case of
Boussinesq convection. In particular, we shall consider the case of a
rotating spherical fluid shell as shown in figure \ref{f.010}(b). 
\blue{Since we are using only the asymptotic results of \textsection 2
which are independent of the} \reviii{radial} {length scale $d$ in the formulation of the problem,  
we are free to interpret $d$ from now on as the thickness of the  
spherical shell. Accordingly the inner and outer radii, $r_i$ and
$r_o$, are given by $\beta/(1-\beta)$ and $1/(1-\beta)$, respectively,
where $\beta$ is defined by $\beta = r_i/r_o$.}  

\citet{Simitev03} have demonstrated that a good
approximation for the onset of convection in rotating spherical shells
can be obtained by solutions of the form (3.8)--(3.10) in B86
found from the model of the rotating annulus with conical ends.
In the spherical case the parameter $\eta^*$ is defined by   
\begin{gather}
\eta^*\equiv\frac{\tau \tan\theta}{r_o \cos \theta} = \frac{\tau x}{r_o^2-x^2}.
\label{eq:14}
\end{gather}
where $\theta$ is the colatitude on the spherical surface with respect
to the axis of rotation and $x=r_o\sin \theta$ represents the distance
from the axis at which convection sets in. 
\reviii{
Although the analysis leading to the results \eqref{eq:11} and
\eqref{eq:12} with $\eta^*_\rho$ replaced by $\eta^*$ as shown in B86
is mathematically rigorous only in the limit of small $\theta$, the
asymptotic expressions compare quite well with the numerical results
at finite angles $\theta$. 
A further improvement may eventually be obtained following
\citet{Calk13} who replaced the two-dimensional procedure by a three-dimensional one.
}

In the presence of density variation the contribution $\eta_\rho^*$ as
defined in the preceding section must be added. Since the density in
the spherical configuration varies not only with distance from the
axis, but also parallel to the axis, an average over the latter
dimension must be taken. The same procedure must be applied to
gravity. 
In order to compare our analytical results with the direct
numerical results of \citet{Jones09}, we shall
use the following explicit expressions defined in \citep{Jones09,Jones11}:
\begin{gather}
\bar \rho \equiv \xi^n \quad \mbox{with} \quad \xi=c_0 + c_1/r,\\
\mbox{and with} \quad c_0= \frac{2\xi_o-\beta -1}{1-\beta},  \quad c_1= \frac{(1+\beta)(1-\xi_o)}{(1-\beta)^2},\nonumber\\
\red{\mbox{where} \quad \xi_i = \frac{\beta + 1 - \xi_o}{\beta} \quad \mbox{and} \quad \xi_o = \frac{\beta + 1}{\beta \exp(N_\rho/n)+1}\nonumber}
\label{eq:15}
\end{gather}
are the values of $\xi$ at the inner and outer boundaries\red{. Here}
$N_\rho$ is the number of density scale heights,
$N_\rho=n\ln(\xi_i/\xi_o)$, and $n$ is the gas polytropic index.

\begin{figure}
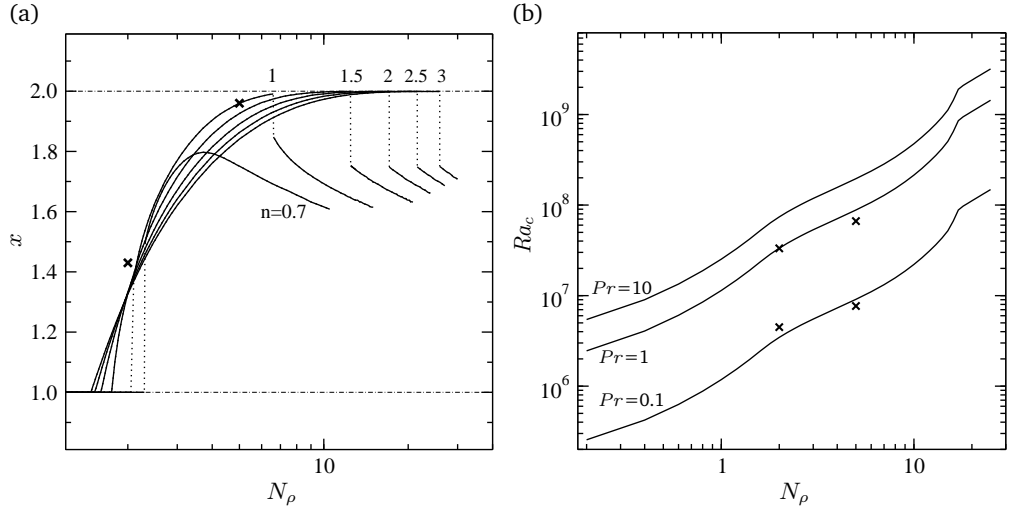

\begin{center}
\psfrag{Nr}{$N_\rho$}
\psfrag{Ra}{$\Rac$}
\psfrag{x}{$x$}
\psfrag{P=1}{\hspace{-1mm}{\scriptsize $\Pr$=1}}
\psfrag{P=10}{\hspace{-2mm}{\scriptsize $\Pr$=10}}
\psfrag{P=0.1}{\hspace{-1mm}{\scriptsize $\Pr$=0.1}}
\psfrag{(a)}{}
\psfrag{(b)}{}
\begin{tabular}{ll}
(a) & (b)\\
\epsfig{file=fig04a.eps,width=0.48\textwidth,clip=}&
\epsfig{file=fig04b.eps,width=0.49\textwidth,clip=}\\
\end{tabular}
\end{center}
\caption{(a) The position of the minimal critical Rayleigh number,
{$\Rac$, as a function of $N_\rho$ for the case $\beta=0.5$, $\Pr=1$,
  $\tau=10^5$  with $n$ as indicated in the plot.}
(b) The minimal critical Rayleigh number, $\Rac$, as a function of $N_\rho$
for  {$\beta=0.5$}, $\tau=10^5$, $n=2$ and $\Pr=0.1,1,10$ as indicated in
the plot. The crosses show values obtained by \citet{Jones09}
from direct numerical solution of the linearised spherical problem
with $n=2$.}  
\label{fig03}
\end{figure}

The definitions \eqref{eq:5} and \eqref{eq:6} thus become modified to
\begin{gather}
\label{eq:16}
\begin{align}
\eta^*_\rho & = -\frac{\tau}{\sqrt{r_o^2-x^2}}
\int\limits_{0}^{\sqrt{r_o^2-x^2}}  \frac{d}{dx}\ln\Big(\bar\rho\big(\sqrt{x^2+z^2}\big)\Big)  dz \nonumber \\
& =\frac{n\tau}{\sqrt{r_o^2-x^2}}\int\limits_{0}^{\sqrt{r_o^2-x^2}}\frac{c_1xdz}{(c_0\sqrt{x^2+z^2} +c_1)(x^2+z^2)}.
\end{align}
\end{gather}
An analytical expression for this integral can be obtained, but it is
lengthy and will not be given here. 
\blue{
The parameters $\eta^*$ and $\eta^*_\rho$ are displayed as
functions of $x$ in figure \ref{fig02a}(a) in the special  case $n=2$,
$\beta=0.5$.
}

In order to apply the two-dimensional approximation to the
three-dimensional formulation used by \citet{Jones09,Jones11}, we must take
into account that the entropy gradient of the purely conducting state
is spatially dependent and that the same holds for the gravity term
which in \citep{Jones09,Jones11} is assumed to vary in proportion to
$r^{-2}$. For the cylindrical approximation only the component
perpendicular to the axis of rotation is
relevant. For this component  in the formulation of 
\citet{Jones09,Jones11}, we get in place of $\R$ in equation \eqref{eq:8}
\begin{gather}
\Ra \frac{x}{r} \frac{\Delta s}{r^2} \left[\frac{n c_1 x}{r^3\red{\xi}^{n+1}}
  \frac{1}{\red{\xi}_o^{-n}-\red{\xi}_i^{-n}}\right],
\label{eq:17}
\end{gather}
where the term inside the square brackets denotes the negative $x$ derivative of the entropy
in the \reviv{motionless} purely conducting state.
The average of expression \eqref{eq:17} over the cylindrical surface intersecting the
spherical shell at the distance $x$ from the axis thus leads to the ratio
between the Rayleigh number $\R$ of the cylindrical model and the
Rayleigh number $\Ra$ introduced by \citet{Jones09,Jones11} in the
spherical case
\begin{gather}
\frac{\R}{\Ra} = \frac{n c_1 x^2}{\sqrt{r_o^2-x^2}}
\int_{0}^{\sqrt{r_o^2-x^2}} \frac{dz}{(x^2+z^2)^3
  (c_0+c_1/\sqrt{x^2+z^2})^{n+1} (\xi_o^{-n}-\xi_i^{-n})}.
\label{eq:18}
\end{gather}
\blue{
The reciprocal of this function which is independent of
$\tau$ and $\Pr$ has been plotted in figure \ref{fig02a}(b) in the
special case $n=2$, $\beta =0.5$.
}

\section{Comparison of the asymptotic results with numerical data}
\label{sec4}
In this section the asymptotic results derived above will be compared
with numerical data found in the literature for the onset of
anelastic convection in rapidly rotating spherical shells. The goal
of this comparison is not to demonstrate an optimal quantitative
agreement, but to show that the asymptotic expressions reflect all
qualitative properties of the numerical results quite well and can
thus be used for explorations of regions of parameter space that
may not be easily accessible to direct numerical integrations. 
\begin{figure}
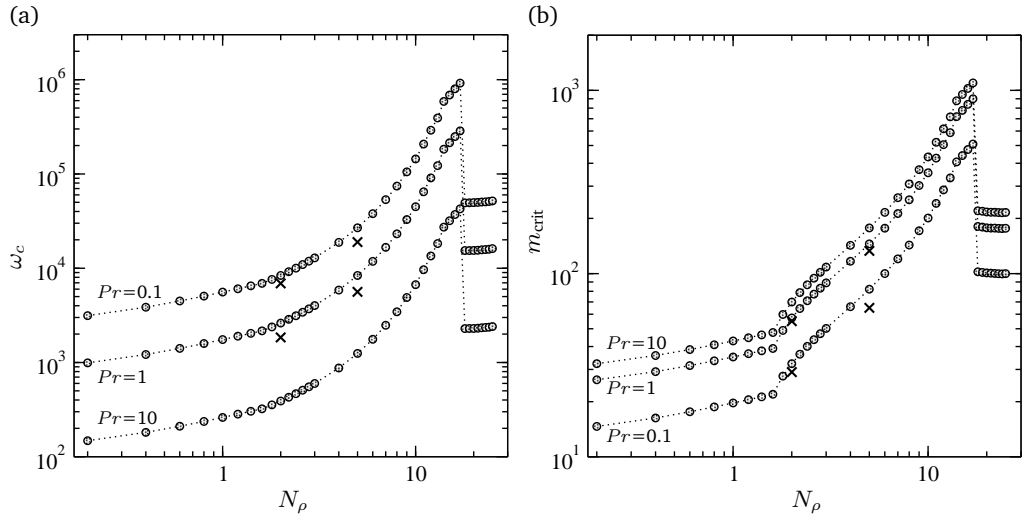

\begin{center}
\psfrag{Nr}{$N_\rho$}
\psfrag{-omg}{$\omega_c$}
\psfrag{m}{$m_\text{crit}$}
\psfrag{P=1}{\hspace{0mm}{\scriptsize $\Pr$=1}}
\psfrag{P=10}{\hspace{0mm}{\scriptsize $\Pr$=10}}
\psfrag{P=0.1}{\hspace{0mm}{\scriptsize $\Pr$=0.1}}
\psfrag{(a)}{}
\psfrag{(b)}{}
\begin{tabular}{ll}
(a) & (b)\\
\epsfig{file=fig05a.eps,width=0.495\textwidth,clip=}&
\epsfig{file=fig05b.eps,width=0.49\textwidth,clip=}\\
\end{tabular}
\end{center}
\caption{(a) The global critical frequency $\omega_c$ and (b) the
critical azimuthal  wave number $m_\text{crit}$ as a function of $N_\rho$
for {$\beta=0.5$}, $\tau=10^5$, $n=2$ and $\Pr=0.1,1,10$  as indicated
in the plot. The crosses show values obtained by \citet{Jones09}
from direct numerical solution of the linearised spherical problem.}
\label{fig04}
\end{figure}

Combining the effect of the density stratification \eqref{eq:16}  and that of
the boundary inclination \eqref{eq:14} at the spherical surface we find 
\begin{gather}
\etaP(x)\equiv\frac{|\eta^*_\rho+\eta^*|\Pr}{\sqrt{2}(1+\Pr)},
\label{eq:19}
\end{gather}
yielding the asymptotic critical values for the azimuthal wavenumber
$\alpc$ and for the Rayleigh number $\Rc(x)$
\begin{gather}
\alpc(x)=\eta^{1/3}_P(x),  \hspace{2cm} \Rc(x)= 3\eta^{4/3}_P(x).
\label{eq:20}
\end{gather}
\blue{
The dependence of $\Rc(x)$ on the fourth power of $\alpc$ demonstrates  
again that it is independent of the chosen  length scale $d$. The  
characteristic length of convection is given by its azimuthal  
wavelength which does not depend on any radial length scale in  
the asymptotic limit.
}
\reviv{This property is also evident from figure 2 of
  \citet{Jones09} in which the radial extent of the convection
  columns does not reflect any connection with the thickness $d$ of the
  spherical fluid shell.} 

\begin{figure}
\begin{center}
\psfrag{Nr}{$N_\rho$}
\psfrag{Ra}{$\Rac$}
\psfrag{x}{$x$}
\psfrag{a}{(a)}
\psfrag{a}{(b)}
\epsfig{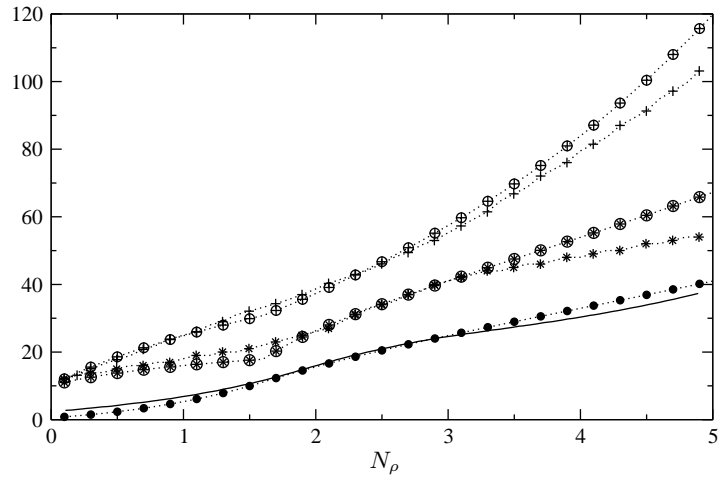}
\end{center}
\blue{\caption{{Comparison with direct numerical results shown in
figure 1(a) of \citet{Jones09}. 
The values $\Rac \times 10^{-5}$, $m_\text{crit}$ and
$\omega_\text{crit} \times 10^{-1}$  reported by \citet{Jones09}
(solid line, stars and plus-signs, respectively) are plotted against
the corresponding values of $\Rac \times 10^{-5}$, $m_\text{crit}$ and
$2/3 \times \omega_c \times 10^{-1}$ obtained from our analysis
(bold dots, circled stars and circled plus-signs, respectively).
The factor 2/3 in front of the values of $\omega_c$ obtained
from expression \eqref{eq:21} has been introduced to facilitate the
comparison of the shapes of the two frequency curves.
The values of the parameters are $\Pr=1$, $\tau=10^4$, $n=2$,
$\beta=0.5$ and $N_\rho$ varies along the abscissa. 
\label{fig05a}
}}}
\end{figure}

The value $\Ra(x)$ is obtained through application of relationship \eqref{eq:18}.
This function is plotted in figure \ref{fig02} for various values of
$N_\rho$ in the case $\beta=0.5, \tau=10^5, n=2$ for the Prandtl
number $\Pr=1$. For $\Pr=0.1$ the same plots are obtained except that the
Rayleigh number is decreased by the factor $(0.2/1.1)^{4/3}$ as
follows from the asymptotic dependence $\big(\Pr/(1+\Pr)\big)^{4/3}$
of the Rayleigh number on the Prandtl number. While 
  the values indicate some quantitative discrepancies with the data displayed
  in \citep{Jones09} the plots show the same qualitative
  features. The most notable feature is that the location of initial
  instability as given by the minimum value of the critical 
  Rayleigh number gradually changes position from near the inner boundary to
  near the outer boundary with increasing $N_\rho$ as shown in
  \citep{Jones09} and also in \citep{Gas12}. However, for values
  larger than $N_\rho=17$ at 
  $\Pr=1$ the location of initial instability abruptly jumps inside the
  fluid layer. This prediction of our analysis has
  not been observed previously as  values of $N_\rho>10$, even though
  physically relevant, are beyond the range of direct numerical
  simulations. 

\blue{As is already evident from figure \ref{fig02a}(a), the dependence of the minimum value of $\Ra$ on $x$ is mainly
governed by  the variation of the density described by
$\eta_\rho^*$. The contribution $\eta^*$ from the inclination of the
outer spherical boundary provides a minor supplement. But since this
latter contribution diverges at the equator, it prevents $x=2$ from being
a location of the onset of convection. Note that a grid step of
$0.001$ in $x$ has been used for the computation of figure \ref{fig02}. 
}

\red{The general dependence of the position of initial instability as a function of
the value of $N_\rho$ is illustrated in figure
\ref{fig03}(a) in the case $n=2$. Analogous curves for different values
of $n$ are also shown there in order to demonstrate the strong
dependence on the polytropic index $n$ of the shift in the onset
location from the outer boundary to the interior.} 
Figure \ref{fig03}(b) shows the minimum critical
Rayleigh number as functions of $N_\rho$ \brown{and figure \ref{fig04} shows
the critical frequency $\omega_c$ and wavenumber, $m_\text{crit}$, as
functions of $N_\rho$.} 
The curves exhibit a
remarkable qualitative agreement with the results shown in figures 1,
3, 5 and 7 of \citet{Jones09} that were obtained by direct numerical
solution of the linearised spherical problem.  
\brown{For instance, in figure \ref{fig05a} the values of $\Rac$, $m_\text{crit}$ and
$\omega_c$ as a function of $N_\rho$ obtained from our model are
compared with the corresponding values as shown in figure 1(a) of
\citet{Jones09} that were obtained by these authors in direct
numerical simulations. While the curves for $\Rac$ and $m_\text{crit}$
compare well, the agreement of the frequencies $\omega_c$ is less
satisfactory. This discrepancy is discussed further below.}
\red{The dependences on other values of $\Pr$ and $\tau$ can easily be inferred from the general relationships
\begin{gather}
\Rac\sim\left(\frac{\tau \Pr}{1+\Pr}\right)^{\frac{4}{3}}, 
\quad 
\omega_c\sim\left(\frac{\tau}{\sqrt{\Pr}(1+\Pr)}\right)^{\frac{2}{3}},
\quad 
m_c\sim\left(\frac{\tau \Pr}{1+\Pr}\right)^{\frac{1}{3}}.
\label{eq:200}
\end{gather}
}
\blue{
The fact that in the above-mentioned figures of \citet{Jones09} the
relatively low value of $\tau=10^4$ has been used may be responsible
for parts of the deviations from the asymptotic results. The
asymptotic dependences \eqref{eq:200} on $\tau$ are indicated in the
case of $\Pr=1$ in figure 6 of \citet{Jones09}, but significant
deviations of the slope of numerical results are still noticeable in
the range $10^4 < \tau < 10^5$.
}

\begin{table}
\caption{Comparison of asymptotic results with direct numerical
  results of \citet{Jones09} in the case $\tau=10^5$, {$\beta=0.5$},
  $N=2$ and   $n=2$.}
\vspace{4mm}
\begin{center}
\begin{tabular}{l@{\extracolsep{2mm}}l@{\extracolsep{10mm}}l@{\extracolsep{2mm}}l@{\extracolsep{6mm}}l@{\extracolsep{2mm}}l}
& & 
\multicolumn{2}{c}{$\Pr=1$ , $x=1.43$} & 
\multicolumn{2}{c}{$\Pr=0.1$, $x=1.6$}  \\[3mm] 
& eqn. &  Asymptotic & Numerical & Asymptotic & Numerical \\\hline
$\eta^\ast$ & \eqref{eq:14}&$3.89 \cdot 10^4$ & & $4.17\cdot 10^4$ & \\
$\eta_\rho^\ast$ & \eqref{eq:16} & $2.40 \cdot 10^5$ & & $2.79\cdot 10^5$ & \\
$\etaP$ & \eqref{eq:19}& $8.482 \cdot 10^4$ & & $1.793\cdot 10^4$ & \\
$\alpc$ & \eqref{eq:20}& $43.9$ & $42.4\quad(38.5)$ & $26.2$ & $22.6\quad(18.1)$\\
$\Rc$ & \eqref{eq:20}& $1.118 \cdot 10^7$ & & $1.408\cdot 10^6$ & \\
$\Rac$ & \eqref{eq:18} & $3.366 \cdot 10^7$ & $3.326\cdot10^7$ & $3.761\cdot 10^6$ & $4.685\cdot10^6$\\
$\omega_c$ & \eqref{eq:21} &$2723$ & $1844$ & $9688$ & $6901$\\
\end{tabular}
\end{center}
\vspace*{3mm}
\label{tab1}
\end{table}

\brown{A direct numerical} comparison with the results shown in figures 2
and 4 of \citet{Jones09} is provided in table \ref{tab1}. Here the
$x$ value of 
the position of the maximum of the amplitude of convection as shown in
figures 2 and 4 \brown{of \citet{Jones09}} has been used in calculating
the various parameters of the asymptotic theory.
In  comparing the critical wavenumbers a difficulty 
arises in that the asymptotic theory assumes that the convection
columns are aligned in the radial direction, such that their
wave-vector is directed in the azimuthal direction. In the numerical
analysis of \citet{Jones09} the convection columns are spiralling
outwards such that the wave-vector acquires a radial component. This
effect is caused by the \blue{radial derivative} of the parameter
$\eta_\rho$ which is not taken into account in the asymptotic
theory. Hence in the table the absolute value of the wave-vector is
indicated as estimated from figures \blue{2 and 4} of \citet{Jones09}. The
azimuthal component of the wave-vector corresponding to the wavenumber
$m$ is indicated in brackets. 
The critical values of the Rayleigh numbers $\R$ and $\Ra$ (as used in
the formulation of \citet{Jones09})  do not depend strongly on the
wavenumber  and they exhibit a fairly close agreement between
asymptotic and numerical values. \brown{The discrepancy in} the
frequencies $\omega_c$ as given by the asymptotic expression
\begin{gather}
\omega_c=(\eta^*_\rho + \eta^*)/\big(\alpc(1+\Pr)\big)
\label{eq:21}
\end{gather}
still persists. This is caused to some extent by the strong dependence
of $\omega_c$ on $\alpc$. \brown{The discrepancy may be reduced by
taking into account the effect of convexly-curved rather than
straight conical sloping endwalls of the annulus. This effect is
considered in \citet{BuOr1986} where it is shown that a perturbation of
opposite sign to that of $\omega_c$ results due to the curvature. 
Although this correction would reduce the discrepancy we have not
calculated it since it can not be expressed in a simple analytical
form.}

\section{Concluding remarks}
\label{sec5}

As should be expected, the comparison of the asymptotic expressions with
the numerical results does not show as good agreement in the anelastic
case as in the Boussinesq case studied by \citet{Simitev03}. On
the other hand, the conceptional value of the approximate asymptotic
theory increases in proportion to the complexities introduced by
anelastic density stratifications.  

In the present paper only the linear local problem of the onset of
convection has been investigated. An extension could be considered in
connection with the spiralling nature of convection which depends on
the second derivative of the density variation in the
$x-$direction. But since an analytical theory for this effect is not
yet available even in the Boussinesq case of the cylindrical annulus
with varying inclination of the conical end surfaces, such an analysis
will be deferred to future research.  
 
The spiralling of the convection columns is an important feature since
it is associated with Reynolds stresses that generate a differential
rotation. Such a mechanism could eventually be described by an
analytical theory based on an expansion in powers of the amplitude of
motion as has been done \reviv{by \citet{Bu83}} in the case of the
Boussinesq version of the problem. 

\blue{Finally the two-dimensional approximate analysis of the linear
problem  of anelastic convection could be improved through a
three-dimensional {multiscale} analysis as has recently
been done by \citet{Calk13} in the  limit of the Boussinesq approximation.}

\vspace{4mm}
\noindent
\textbf{Acknowledgements}\\
This research has been supported by NASA grant NNX-09AJ85G. 
We acknowledge the hospitality of Stanford University and UCLA.
R.D.S.\ enjoyed a period of study leave granted by the University of
Glasgow and the support of the Leverhulme Trust via Research Project
Grant RPG-2012-600.


\end{document}